# A Missing Hallmark of Cancer: Dysregulation of Differentiation

Zachary Compton[1,]*, Kathleen Hanlon[2,]*, Carolyn C. Compton[3], Athena Aktipis [1,4] †, Carlo C. Maley[1,3] †

* co-first authors

† co-senior authors

[1] Arizona Cancer Evolution Center, Biodesign Institute and School of Life Sciences, Arizona State University, Tempe, AZ

[2] University of Arizona Medical School, Phoenix, AZ

[3] School of Life Sciences, Arizona State University, Tempe, AZ

[4] Department of Psychology, Arizona State University, Tempe, AZ

**Abstract**

Cancer cells possess a nearly universal set of characteristics termed the hallmarks of cancer, including replicative immortality and resisting cell death. Dysregulated differentiation is present in virtually all cancers yet has not yet been described as a cancer hallmark. Like other hallmarks, dysregulated differentiation involves a breakdown of the cellular cooperation that typically makes multicellularity possible - in this case disrupting the division of labor among the cells of a body. At the time that the original hallmarks of cancer were described, it was not known that dysregulated differentiation was mechanistically distinct from growth inhibition, but now that this is known, it is a further reason to consider dysregulated differentiation a hallmark of cancer. Dysregulated differentiation also has clinical utility, as it forms the basis of pathological grading, predicts clinical outcomes, and is a viable target for therapies aimed at inducing differentiation. Here we argue that hallmarks of cancer should be near universal, mechanistically distinct, and have clinical utility for prognosis and/or therapy. Dysregulated differentiation meets all of these criteria.

**Introduction**

The identification of the hallmarks of cancer has been one of the most helpful and influential contributions to understanding cancer, because it brings simplicity, consistency and coherence to the otherwise overwhelming complexity of cancers[1,2]. While cancer genomics has shown that each cancer is a unique mosaic of diverse genetic clones, evolutionary theory helps us understand why this diversity often converges on strikingly similar phenotypes represented by the hallmarks[3]. We can view the hallmarks of cancer as the characteristics that are common across cancers, evolving consistently and independently in each cancer, because they confer a fitness benefit to the neoplastic cells over the surrounding normal somatic cells[3]. All complex multicellular organisms require cooperation between their individual somatic cells[4]. Although complex multicellularity has evolved at least seven times[5], there are five forms of cooperation upon which all multicellular organisms have converged upon: suppression of cell proliferation, controlled cell death, resource allocation, maintenance of the extracellular environment, and division of labor among the somatic cells[4]. Cancer, as a more general problem for multicellularity, can be understood as cells that cheat on the forms of cooperation necessary for building and maintaining a multicellular entitybody (Figure 1). All the current hallmarks of cancer map onto the five foundations of multicellularity, with one exception: there is no hallmark that corresponds to cheating on the division of labor among cells[4]. Here we suggest that there should be an additional hallmark of cancer which corresponds to this breakdown in division of labor. A breakdown of division of labor among cells would manifest as cells not adopting the proper cell types that are necessary for the proper functioning of the organism, i.e., dysregulated differentiation.

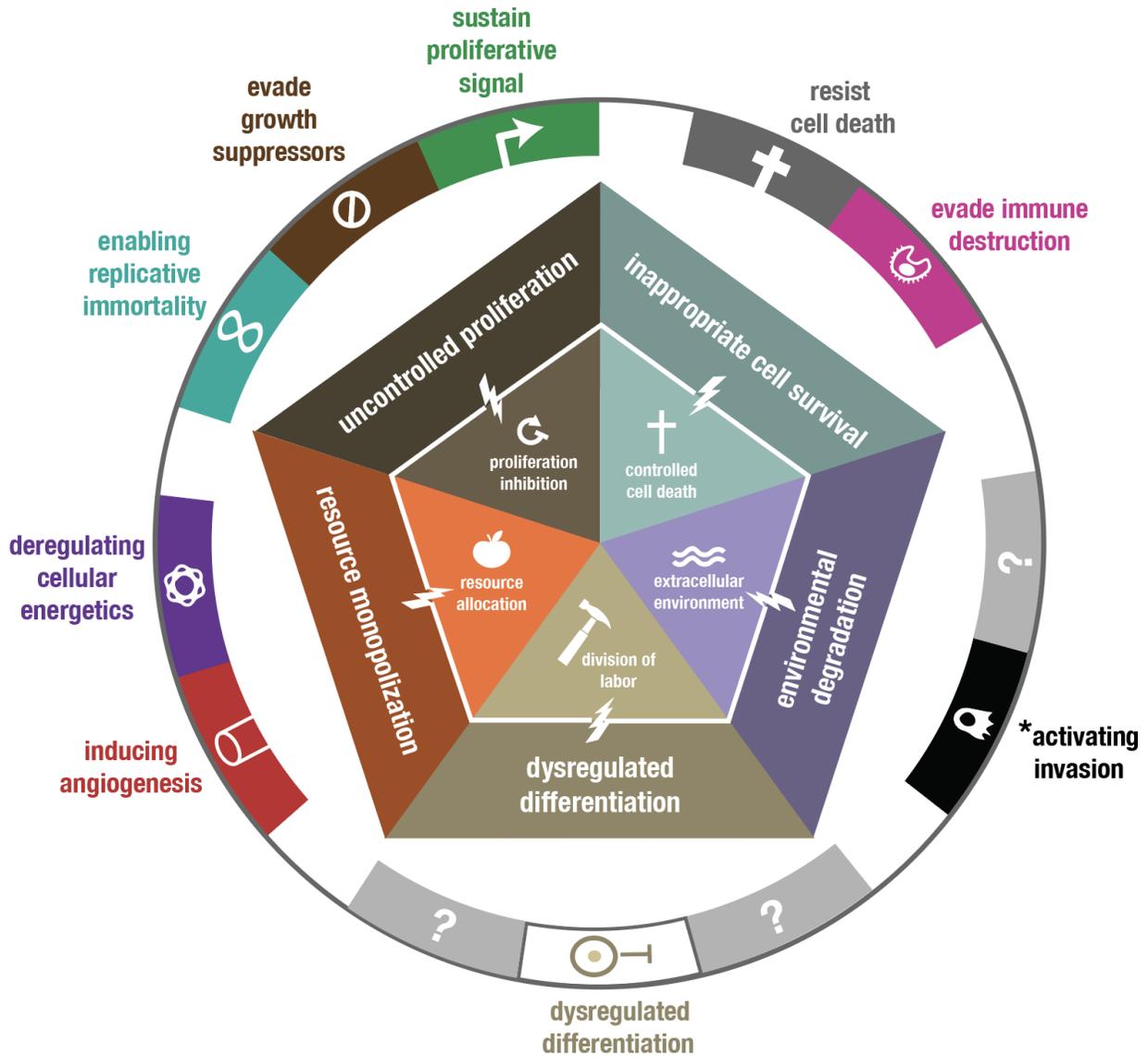

**Figure 1.** Cancer represents a breakdown of the foundations of multicellular cooperation that are necessary for multicellularity to succeed. The breakdown of every foundation of multicellularity corresponds to one or more of the existing hallmarks of cancer, with the exception of division of labor. Adding dysregulated differentiation as an additional hallmark of cancer fills this gap, corresponding to a breakdown in division of labor. There may well be other missing hallmarks, represented here as other gaps in the periphery.

Not only does dysregulated differentiation fill this gap, it also is already a well-recognized universal feature of cancer that is mechanistically distinct from other hallmarks, important for prognosis and a promising target for therapy. As we will argue in this perspective, the cancer hallmarks should not only be universal across cancers, but they should also be mechanistically distinct from one another, as well as diagnostically and therapeutically useful (Table 1). By these criteria, dysregulation of differentiation should be considered a hallmark of cancer.

| **Cancer Hallmark** | **Mechanistically Distinct** | **Diagnostically Functional** | **Therapeutically Relevant** |
|---|---|---|---|
| Sustained Proliferative Signaling | Constitutive activation of proliferative pathways[6] | Proliferative markers such as Ki-67 have been long used in staging/grading cancers[7] | Numerous compounds have demonstrated efficacy against known proliferative pathways[8] |
| Evade Growth Suppressors | Tumor suppressor pathways cannot be fully functional in metastatic disease[9] | Although characterized in childhood retinoblastoma, the RB pathway is mutated the majority of human cancers[10] | RB mutation status can significantly guide the clinical management of a variety of cancer types[10] |
| Avoid Immune Destruction | Through the expression of self antigens and manipulation via the tumor microenvironment, many tumor cells escape immune destruction[11] | Intratumor leukocyte infiltration can be used as a prognostic index determining anti-tumor immune activity[12] | There a several therapeutic targets such as PD-1, PD-L1, CTLA4, and Th1 that can potentially counter immune evasion[13,14] |
| Enable Replicative Immortality | Cancer cells are able to restore and maintain telomere functionality[15] | Telomerase activity provides insight to tumor differentiation status[16] | Targeting cyclin dependent kinases such as PI3K could trigger cancer cell senescence[17]. Telomerase is a target for cancer therapy[18] |
| Activate Invasion & Metastasis | The ability of cancer cells to penetrate the basement membrane and disseminate into different tissues[19,20] | Circulating tumor cells (CTCs) can be assayed for early detection of metastatic disease[21] | Cell adhesion pathways can be targeted for therapy[22] |
| Induce Angiogenesis | Tumors cannot grow beyond 1-2mm$^3$ without establishing their own vasculature[23] | Density of tumor supporting vasculature a useful prognostic tool[24] | Anti-angiogenesis therapies target tumor resource delivery[25] |
| Resist Cell Death | Disruption in the Bcl-2 signaling pathway precludes apoptotic response to DNA damage[26] | Determining the activity of BLC2 family proteins can reveal cells' ability to resist apoptosis[27,28] | Targeting death receptor ligands can trigger tumor cell death[29,30] |
| Deregulate Cellular Energetics | Cancer cells forgo oxidative phosphorylation, relying almost exclusively on glycolysis[31] | Cancer cell metabolic phenotype predicts disease progression[32] | Recognized metabolic alterations are emerging as therapeutic targets[33] |
| **Proposed new hallmark** | **Mechanistically Distinct** | **Diagnostically Functional** | **Therapeutically Relevant** |
| Dysregulated Differentiation | Tumor genomic profiles outline key genetic lesions that grant cancer cells their stem cell qualities[34] | Differentiation is the foundation of tumor grading[35] | Differentiation therapy provides a unique therapeutic target with minimal toxicity[36] |

**Table 1.** Every existing hallmark of cancer has the properties of being mechanistically distinct, diagnostically functional (providing information, either at the genetic, cellular, or tissue level that can be utilized by a physician in diagnosis or prognosis) and therapeutically relevant (providing identifiable targets for therapeutic intervention). In addition to being universal across cancers, dysregulated differentiation exhibits these three features as well, suggesting that it should be included as a hallmark of cancer.

**A universal feature of cancer**

Dysregulation of differentiation is a universal feature of cancers[37,38]. Both genomic and histological evidence indicate that dysregulated differentiation is pervasive (Table 2). Cancers are generally diagnosed by histological features, detectable under a light microscope, that indicate that something has gone wrong in differentiation. Histological examination of differentiation status is a foundational method in the cancer grading system which has long been the cornerstone determining patient prognosis[39,40]. These histological aberrations of differentiation are far ranging, including glands are that improperly formed or are missing altogether. Sometimes there is loss of regulation over a progenitor cell population, that has not fully differentiated, such that it expands to a pathological level, as occurs in most of the hematopoetic neoplasms[37] as well as the undifferentiated clonal expansions in carcinomas. In fact, the generation of a new mass, a neoplasm, is probably impossible as long as differentiation is being properly regulated. Differentiation regulates the proper proportions and number of different cell types in every tissue. The epithelial-to-mesenchymal transition (EMT) common to many cancers is a further example of aberrant differentiation[41,42].

| Cancer Type | Genomic Evidence of Dysregulated Differentiation | Histological Evidence of Dysregulated Differentiation |
|---|---|---|
| Breast | Down regulation of Gata-3 precludes healthy gland differentiation and disrupts luminal cell fate.[43–46] | Tumor differentiation status defines grading scale and strongly predicts patient prognosis[35,47,48] |
| Colorectal | NDRG2 is expressed at low or undetectable levels in high risk/poor prognosis colorectal adenomas[49,50] | Differentiation status of a tumor was more predictive of prognosis than invasive margin and DNA ploidy[51] |
| Prostate | FOXA1 suppression in prostate carcinoma indicative of irregular differentiation patterns[52,53,54] | Lack of full differentiation in prostate cancer precludes the usefulness of serum prostate specific antigen in measuring tumor burden.[39,55] |
| Lung | TRPC channel disruption signals stemcell-like differentiation status[56–58] | Differentiation status is an independent predictor of prognosis in non-small cell lung cancer[59,60] |
| Thyroid | Suppression of Notch signaling mediated differentiation [61–63] | Diversity of thyroid carcinoma subtype founded largely on morphological differentiation[64–66] |
| Bladder | Renewal of Hedgehog signaling pathway can illicit differentiation factors that improve prognosis [67,68] | Tumor cells reveal morphological indications of dysegulated differentiation before chromosomal aberrations[69,70] |
| Stomach | Amplification of Notch1 intracellular domain maintains population of undifferentiated or poorly differentiated cells in carcinoma of the stomach[71–73] | Even well differentiated gastric carcinoma show histological evidence of disruption[74] |
| Cervical | Expression of FOXC2 in cervical tissue correlates with increases in number of poorly differentiated cells[75,76] | Disruption of healthy differentiation can be detected with light microscope and/or positron emission tomography[77] |
| Non-hodgkin Lymphoma | Expression profiles show T-cell differentiation in B-NHL is skewed towards early stages [78] | Phenotypic classification of tumor cells by degree of differentiation informs prognosis[79,80] |
| Endometrial | Karyotypic aberration patterns correlate with histological differentiation[81] | Tissue specific differentiation and hormone receptor positivity are key prognostic factors[82–84] |

| Leukemia | Pax5 loss and t(15;17) translocations both cause differentiation blocks in leukemias[85] | A review of differentiation therapy for leukemia[36,86] Undifferentiated leukemia by light microscopy with myeloid features[36,86,87] |
|---|---|---|
| Kidney | Positive correlation between low PTEN expression and poorer differentiation[88] | Differentiation level by subtype predicts patient outcome[89,90] |
| Melanoma of the skin | Melanoma differentiation associated gene‐7 (MDA7) expression is downregulated in advanced melanoma and virtually undetectable in metastatic disease[91] | Differentiation status and like-ness with other skin markings provides a baseline understanding of disease state |
| Lip, oral cavity | Absence of epithelial keratins defines a de-differentiated state in oral carcinomas[92,93] | Morphological differentiation status, although particularly subjective in the oral cavity, still associated with patient outcome[94,95] |
| Brain and Central Nervous System | Reactivation of Wnt signaling induce neural differentiation and cancer cell death [96–99] | Glioblastoma stem-like cells can hijack differentiation pathways to recruit vascularization[100,101] |
| Ovary | Notch1 overexpression increases with decreasing extent of fully differentiated cells[102–105] | Extent of morphologically poorly differentiated cells within ovarian tumor predicts prognosis[82,106,107] |
| Liver | MYC inactivation in an animal model of HCC induced differentiation and sustained regression of the tumor[108,] Increased LEF1 expression in hepatocellular cancer is associated with poor cellular differentiation and worse prognosis, and | Well differentiated hepatocellular carcinoma presents atypically and yet retains histological evidence of differentiation abrogations[110,111] |

| | regulates tumor differentiation through activation of NOTCH signaling pathways[109] | |
|---|---|---|
| Esophagus | 22% of esophageal squamous cell carcinomas have mutations in genes that regulate esophageal squamous cell differentiation (NOTCH1, NOTCH2 or NOTCH3) [112]<br>In squamous cell carcinoma, Notch3 is repressed by TGF*B*, which blocks terminal differentiation and leads to Notch1 mediated EMT [113] | Majority of esophageal carcinoma shows moderate to completely undifferentiated cell morphology[114] |
| Larynx | Cyclin E overexpression in a majority of laryngeal carcinomas is a key driver of poorly differentiated tumors[115]. | Lymphoepithelioma is an undifferentiated carcinoma of the nasopharyngeal type with propensity for metastasis [116–118] |
| Multiple myeloma | Maintained B cell expression of CD38 perpetuates a sub-differentiated population of cells clonally related to the multiple myeloma plasma cells[119,120] | Morphological indications of plasma cell differentiation level significantly predict clinical outcome[121,122] |

**Table 2.** Evidence for disruption of differentiation, both genetic and histopathological, in the 20 most prevalent types of cancers worldwide[123].

**Convergent somatic evolution**

Cells that stop devoting resources to the tasks inherent to that of their normal differentiated state, and instead devote those resources to proliferation and survival, will have a fitness advantage over cells that continue to devote resources to the specific tasks of their tissue type. Dysregulation of differentiation evolves independently in each cancer because it provides a selective advantage to those cells.

Differentiation is beneficial for organisms because it not only allows for the division of cellular labor, but also because it can lower cancer risk through reducing ongoing cell proliferation[124]. This appears to be one of the mechanisms that organisms have evolved to prevent somatic mutations and the expansion of clones that acquire selective advantages from those mutations[125]. In fact, there are many features of differentiated tissue architecture that function to constrain would-be clonal expansions. In intestinal crypts, which have a high rate of cell turnover, this function is performed by basal apical polarity axis maintained through a basement membrane attachment requirement, apical tight junctions between adjacent cells, and basal hemidesmosomal attachment complexeses[126–130]. Consequently, neoplastic cells gain a cell-level fitness advantage by evading those constraints[3]. The suspension of proliferative abilities in fully differentiated cells is one of the major mechanisms of somatic-level evolutionary suppression, in other words it is a cancer resistance mechanism. However, this also means that there is strong selective pressure on neoplastic cells to evolve the ability to evade full differentiation.

Cairns first pointed out in 1975 that if mutations are gained in transit amplifying cells, which are only partially differentiated, these cells will quickly be flushed from the body with little chance to accumulate additional mutations necessary to cause cancer[125] (Figure 2). In this way, differentiation in tissues with high cell turnover acts as a tumor suppressor. Follow-up mathematical and computational models have shown that alterations in differentiation are likely some of the most universal early lesions in neoplastic progression[131,132].

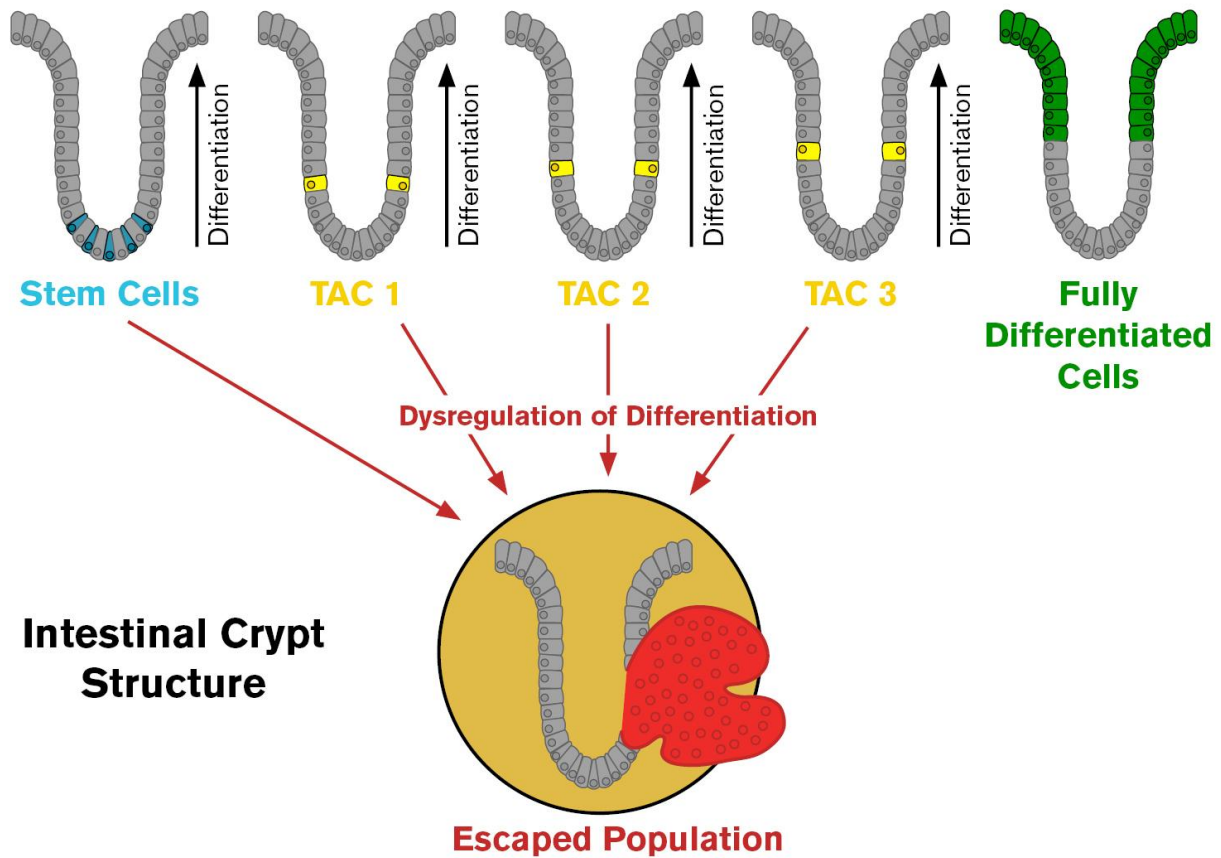

**Figure 2.** In order for a neoplasm to grow, cells must somehow evade the inexorable conveyor belt of differentiation that transforms stem cells into progressive stages of transit amplifying cells, eventually becoming fully differentiated cells, and finally exiting the tissue by apoptosis. There are two ways a neoplasm may form: **a.)** A clone of stem cells may stop producing transit amplifying cells, only dividing symmetrically to produce daughter stem cells. That clone will have a competitive advantage over any stem cell clones that continue to use some of their resources to produce transit amplifying cells. This may be due to an abrogation in the clone's differentiation pathways or through the gaining of independence from stem cell niche signals that would otherwise be required to maintain the stem cell state. **b.)** Alternatively, transit amplifying cells may stop differentiation and so effectively step off of the conveyor belt of differentiation. This gives the non-differentiating (and thus self-renewing) transit amplifying cells a competitive advantage over transit amplifying cells that continue to differentiate.

Both stem cells and transit amplifying cells gain fitness advantages from disrupting differentiation[133]. However, there are generally many more transit amplifying cells than stem cells and so some mathematical models predict that most cancers derive from transit amplifying cells, even if that requires additional mutations to disrupt differentiation[132]. In order to become cancerous, transit amplifying cells must avoid the fate of being sloughed from the off of a proliferating tissue (Figure 2b). Any stem cell that disrupts differentiation and divides symmetrically, producing two daughter stem cells, will have a fitness advantage over other stem cells that divide asymmetrically and use some of their resources to produce non-stem cells (Figure 2a). In summary, there are good evolutionary reasons to expect that virtually all cancer cells can gain a fitness benefit from disrupting differentiation, which explains why dysregulation of differentiation consistently evolves and is a universal feature of cancers.

**Mechanisms of dysregulation**

In their original hallmarks paper, Hanahan and Weinberg discuss the apparent strategy of tumor cells to promote growth by avoiding terminal differentiation. The authors specifically cite the Mad-Max complex, the inactivation of APC/B-catenin pathway in colon carcinogenesis, and the erbA oncogene in avian erythroblastosis[2]. At the time, the dysregulation of differentiation was not included in the hallmarks of cancer because the mechanisms of differentiation could not be distinguished from an insensitivity to antigrowth signals and limitless replicative potential. It was not clear whether loss of differentiation was simply a loss of growth inhibition or an independent factor in carcinogenesis.

In general, differentiation and growth inhibition are tightly, and mechanistically coordinated. However, there are instructive cases of fully differentiated cells that are still proliferative, including beta cells in the pancreas, hepatocytes in the liver, T-cells, and fibroblasts in numerous tissue types[134–140]. These exceptions show that there is a fundamental distinction between loss of proliferative ability and differentiation, though they co-occur often.

There has been ample documentation of interruptions in key differentiation pathways, separate from growth inhibition pathways, that are conserved across cancer types, ultimately preventing true terminal differentiation. Notch signaling plays a complex, and not fully understood, role in distinct differentiation signaling pathways[141]. In T-cell acute lymphoblastic leukemia (T-ALL) chromosomal translocations result in constitutive *Notch1* signaling that precludes terminal differentiation[142–148]. Rangarajan and colleagues demonstrated that *Notch1* deletion in keratinocytes resulted in hyperplasia and generalized dysregulation of known differentiation markers[149]. In addition, constitutive expression of the MYC oncogene is common in human cancers and has a well-established role in prevention of differentiation and sustaining proliferative signals[150–157]. For instance, in a murine model of liver cancer inactivation of MYC was sufficient to differentiate the tumor into normal hepatocytes[108]. Interestingly, c-myc expression drives the differentiation of keratinocytes where *Notch1* appears to play the role of a tumor suppressor[158–161].

The identification of differentiation specific pathway alterations holds the potential to serve as an indicator for therapeutic response. In the crypt structures of the intestinal epithelium, progenitor stem cells are characterized by high expression of Leucine-rich repeat-containing G-protein coupled receptor (LGR5)[162]. Similarly high levels of expression are seen in colorectal cancers, where it is indicative of catastrophic Wnt/β-catenin signaling deregulation and worse patient outcomes[162–164].

**Well-differentiated tumors**

Well-differentiated cancers sometimes can be difficult to distinguish from reactive changes in tissue, such as hyperplasia, or benign tumors by light microscopy. However, even if indiscernible by visual inspection, molecular evidence shows the presence of dysregulated differentiation in well-differentiated tumors. Well-differentiated tumors exhibit a less differentiated molecular profile with elevated expression of precursor genes and lower expression of tissue specific genes compared to healthy tissue[165]. Gene expression studies in histologically differentiated thyroid cancers have found a disruption in differentiation on a molecular level as compared to benign thyroid tissue. Using primary thyroid cancers, Yu *et al* demonstrated that Notch-1 expression was downregulated in differentiated thyroid cancer tissues compared to benign thyroid tissues and that decreased Notch-1 expression was associated with more aggressive tumors with extrathyroidal invasion[166]. Restoration of Notch-1 expression in a metastatic, differentiated thyroid carcinoma cell line led to a reduction in cell growth and tumor cell migration[166]. The Cancer Genome Atlas Research Network investigated the relationship between driver mutations BRAFV600E and RAS and differentiation in papillary thyroid cancer, a typically well-differentiated cancer by histology[167]. Differentiation of over 350 PTCs was quantified and scored by measuring mRNA expression of 16 thyroid function genes, with a lower score indicating decreased differentiation[167]. Interestingly, increased differentiation scores were correlated with the PTC driver mutation BRAFV600E, while decreased differentiation scores were correlated with the driver mutation RAS[167]. Further, upon pathological examination, tumors with lower differentiation scores by mRNA expression were found to have subtle architectural changes that generated more poorly formed and complex papillary structure with fewer follicles[167]. These findings suggested that certain driver mutations may contribute to decreased differentiation in thyroid cancer[167]. As the new molecular tools under development are advanced for clinical application, the feasibility of identifying lack of terminal differentiation even in the most well-differentiated cancers increases.

**Prognostic importance**

Poorly differentiated cancer cells are known to be much more aggressive than their well differentiated counterparts, a fact which plays a critical role in predicting patient outcome[39,74,168,169]. Well-trained pathologists have long been able to accurately assign patient prognosis through tumor grade although the process is heavily burdened with the inherent subjectivity in assessing differentiation optically and the morphological variation that is inherent in most tumors. Advances in cancer genomics have validated genetic attributes that resemble stem cells[170–172] as prognostic indicators. Similar to histopathological grading systems, differentiation gene-expression profiles can predict patient outcomes[170–172]. A 2017 study examined the global gene expression profile of cancer cells and stratified them based on their distance in expression from that of stem cells to fully differentiated cells, using several different histologies including carcinomas, sarcomas, and hematologic malignancies[173]. This methodology allowed for the derivation of a novel cancer gene expression signature found in all undifferentiated forms of the diverse cancers studied. For all subtypes analyzed, tumors most similar in expression to stem cells were both histologically less differentiated and clinically more aggressive. Furthermore, they also demonstrated that where a cancer fell on this spectrum predicted the patient's survival. Work by Riester et al.[173] and others has shown that there are objective measures of cellular differentiation, utilizing descriptive genetic profiles that detail where on a spectrum from "stemness" to full differentiation a given cancer cell lies. Grading with molecular assays that measure the hallmarks of cancer enriches our ability to make clinical predictions while introducing novel quantification of differentiation status through genetic analysis.

**Promising clinical opportunities**

Differentiation should be considered a hallmark of cancer not only due to its universality and distinct cellular mechanisms that drive cancer, but also because the biological mechanisms can be targeted by available therapies that have already shown promise in the clinic. Rather than killing both healthy and tumor cells as do typical chemotherapeutics, differentiation therapies capitalize on the ability of cytokines to promote terminal differentiation of tumor cells and halt their capacity to self-renew[174–177]. This option is especially promising for patients suffering from comorbidities who are unable to receive high-dose chemotherapy due to its significant toxicity.

The first successful clinical application of differentiation therapy was the use of All-*trans* Retinoic Acid (ATRA) for acute promyelocytic leukemia (APL). ATRA induces APL blasts to terminally differentiate[36]. The current standard of care for treatment of APL involves the combination of ATRA and arsenic, making APL now a highly curable disease with 5-year disease-free survival rates that exceed 90%[178].

Outside of APL, differentiation therapy has been gaining traction in the treatment of acute myeloid leukemia (AML)[36,179–182]. A recent preclinical study has identified a novel, highly potent and selective inhibitor that induces differentiation *in vitro* and *in vivo* by inhibiting dihydroorotate dehydrogenase across multiple AML subtypes[179]. A promising phase-1 trial of this inhibitor, BAY 2402234, is currently ongoing for myeloid malignancies (NCT 03404726).

Whether differentiation therapy shows similar effects in cancers apart from the hematological malignancies is worth investigating. Cancer stem cells (CSCs), also known as tumor-initiating cells, represent one such target for differentiation therapy[183–189]. First identified in AML in 1997, they have since been identified in brain cancer, colon cancer, pancreatic cancer, prostate cancer, melanoma, and more[190,191]. They are highly resistant to traditional chemotherapy and radiotherapy, which may be due in part to their relative slow growth and high expression of anti-apoptotic proteins[192]. Differentiation therapy is a promising tactic that may induce these CSCs into non-stem cancer cells with limited self-renewal potential. These non-stem cells could possibly be better targeted by conventional therapies[193,194].

Another logical application of differentiation therapy would be in tumors that are collectively known as "blastomas" or small round blue cell tumors, named after their histological appearance, which is monotonous and characterized by lack of differentiation features. These relatively undifferentiated tumors originate from stem cell progenitors and occur almost exclusively in pediatric patients. Neuroblastoma is one of the small round blue cells tumors. It is famous for frequent spontaneous regression or differentiation into a benign ganglioneuroma[195]. Recent evidence shows that neuroblastomas are composed of cells from two super-enhancer associated differentiation states: undifferentiated mesenchymal cells and committed adrenergic cells[196]. Nevertheless, cells from either state can interconvert, highlighting a potential mechanism of tumor relapse as mesenchymal cells are known to be relatively resistant to chemotherapy[196]. Furthermore, these preserved differentiation pathways have been successfully targeted in vitro with retinoids[197,198], a response categorized with cell proliferation arrest and a markedly lower MYCN expression[197].

Differentiation therapy can only work if some differentiation pathways remain intact in a cancer and can be stimulated by an intervention. Due to natural selection at the somatic level for the dysregulation of differentiation, it may not always be possible to induce differentiation.

**Conclusions**
Dysregulation of differentiation is a universal phenotype, found in virtually all cancers (Table 2). The degree of differentiation has long been used in oncology for diagnosis as well as prognosis, and advances in genomic analyses have shown promise for improving prognosis. Dysregulation of differentiation is molecularly distinct from the other hallmarks, including evading growth suppressors, and it has been successfully targeted for therapy in acute promyelocytic leukemia and neuroblastoma. Further, it is clear that dysregulated differentiation is a breakdown of multicellular cooperation, and the only aspect of this breakdown of multicellular cooperation that is not already represented in the hallmarks of cancer[4]. Together, this suggests that dysregulated differentiation is a missing hallmark that should be added to the commonly accepted list of shared phenotypes of cancer.

## Acknowledgments

This work was supported in part by NIH grants U54 CA217376, U2C CA233254, P01 CA91955, R01 CA170595, R01 CA185138 and R01 CA140657 as well as CDMRP Breast Cancer Research Program Award BC132057, the Arizona Biomedical Research Commission grant ADHS18-198847, the Wissenshaftskollege zu Berlin and the President's office at ASU. The findings, opinions and recommendations expressed here are those of the authors and not necessarily those of the universities where the research was performed or the funding agencies.

## References:

1. Hanahan, D. & Weinberg, R. A. Hallmarks of cancer: the next generation. *Cell* **144**, 646–674 (2011).

2. Hanahan, D. & Weinberg, R. A. The hallmarks of cancer. *Cell* **100**, 57–70 (2000).

3. Fortunato, A. *et al.* Natural Selection in Cancer Biology: From Molecular Snowflakes to Trait Hallmarks. *Cold Spring Harb. Perspect. Med.* **7**, (2017).

4. Aktipis, C. A. *et al.* Cancer across the tree of life: cooperation and cheating in multicellularity. *Philos. Trans. R. Soc. Lond. B Biol. Sci.* **370**, (2015).

5. Knoll, A. H. The multiple origins of complex multicellularity. *Annu. Rev. Earth Planet. Sci.* **39**, 217–239 (2011).

6. Evan, G. I. & Vousden, K. H. Proliferation, cell cycle and apoptosis in cancer. *Nature* **411**, 342–348 (2001).

7. Gerdes, J. Ki-67 and other proliferation markers useful for immunohistological diagnostic and prognostic evaluations in human malignancies. *Semin. Cancer Biol.* **1**, 199–206 (1990).

8. Feitelson, M. A. *et al.* Sustained proliferation in cancer: Mechanisms and novel therapeutic targets. *Semin. Cancer Biol.* **35 Suppl**, S25–S54 (2015).

9. Amin, A. R. M. R. *et al.* Evasion of anti-growth signaling: A key step in tumorigenesis and potential target for treatment and prophylaxis by natural compounds. *Semin. Cancer Biol.* **35 Suppl**, S55–S77

(2015).

10. Du, W. & Searle, J. S. The rb pathway and cancer therapeutics. *Curr. Drug Targets* **10**, 581–589 (2009).

11. Finn, O. J. Immuno-oncology: understanding the function and dysfunction of the immune system in cancer. *Ann. Oncol.* **23 Suppl 8**, viii6–9 (2012).

12. Fridman, W. H. *et al.* Prognostic and predictive impact of intra- and peritumoral immune infiltrates. *Cancer Res.* **71**, 5601–5605 (2011).

13. Vinay, D. S. *et al.* Immune evasion in cancer: Mechanistic basis and therapeutic strategies. *Semin. Cancer Biol.* **35 Suppl**, S185–S198 (2015).

14. Ribas, A. & Wolchok, J. D. Cancer immunotherapy using checkpoint blockade. *Science* **359**, 1350–1355 (2018).

15. Hahn, W. C. & Meyerson, M. Telomerase activation, cellular immortalization and cancer. *Ann. Med.* **33**, 123–129 (2001).

16. Kim, N. W. Clinical implications of telomerase in cancer. *Eur. J. Cancer* **33**, 781–786 (1997).

17. Yaswen, P. *et al.* Therapeutic targeting of replicative immortality. *Semin. Cancer Biol.* **35 Suppl**, S104–S128 (2015).

18. Lee, H.-S. *et al.* Systematic Analysis of Compounds Specifically Targeting Telomeres and Telomerase for Clinical Implications in Cancer Therapy. *Cancer Res.* **78**, 6282–6296 (2018).

19. Pachmayr, E., Treese, C. & Stein, U. Underlying Mechanisms for Distant Metastasis - Molecular Biology. *Visc Med* **33**, 11–20 (2017).

20. Gupta, G. P. & Massagué, J. Cancer metastasis: building a framework. *Cell* **127**, 679–695 (2006).

21. Maheswaran, S. & Haber, D. A. Circulating tumor cells: a window into cancer biology and metastasis. *Curr. Opin. Genet. Dev.* **20**, 96–99 (2010).

22. Li, D.-M. & Feng, Y.-M. Signaling mechanism of cell adhesion molecules in breast cancer metastasis: potential therapeutic targets. *Breast Cancer Res. Treat.* **128**, 7–21 (2011).

23. Carmeliet, P. & Jain, R. K. Angiogenesis in cancer and other diseases. *Nature* **407**, 249–257 (2000).

24. Weidner, N. Intratumor microvessel density as a prognostic factor in cancer. *Am. J. Pathol.* **147**, 9–19 (1995).
25. Cherrington, J. M., Strawn, L. M. & Shawver, L. K. New paradigms for the treatment of cancer: the role of anti-angiogenesis agents. *Adv. Cancer Res.* **79**, 1–38 (2000).
26. Adams, J. M. & Cory, S. The Bcl-2 apoptotic switch in cancer development and therapy. *Oncogene* **26**, 1324–1337 (2007).
27. Letai, A. G. Diagnosing and exploiting cancer's addiction to blocks in apoptosis. *Nat. Rev. Cancer* **8**, 121–132 (2008).
28. Glinsky, G. V. & Glinsky, V. V. Apoptosis amd metastasis: a superior resistance of metastatic cancer cells to programmed cell death. *Cancer Lett.* **101**, 43–51 (1996).
29. Fisher, D. E. Apoptosis in cancer therapy: crossing the threshold. *Cell* **78**, 539–542 (1994).
30. Kelley, S. K. & Ashkenazi, A. Targeting death receptors in cancer with Apo2L/TRAIL. *Curr. Opin. Pharmacol.* **4**, 333–339 (2004).
31. Cairns, R. A., Harris, I. S. & Mak, T. W. Regulation of cancer cell metabolism. *Nat. Rev. Cancer* **11**, 85–95 (2011).
32. Isidoro, A. *et al.* Breast carcinomas fulfill the Warburg hypothesis and provide metabolic markers of cancer prognosis. *Carcinogenesis* **26**, 2095–2104 (2005).
33. Teicher, B. A., Linehan, W. M. & Helman, L. J. Targeting cancer metabolism. *Clin. Cancer Res.* **18**, 5537–5545 (2012).
34. Ben-Porath, I. *et al.* An embryonic stem cell-like gene expression signature in poorly differentiated aggressive human tumors. *Nat. Genet.* **40**, 499–507 (2008).
35. Elston, C. W. & Ellis, I. O. Pathological prognostic factors in breast cancer. I. The value of histological grade in breast cancer: experience from a large study with long-term follow-up. *Histopathology* **19**, 403–410 (1991).
36. Nowak, D., Stewart, D. & Koeffler, H. P. Differentiation therapy of leukemia: 3 decades of development. *Blood* **113**, 3655–3665 (2009).

37. Tenen, D. G. Disruption of differentiation in human cancer: AML shows the way. *Nat. Rev. Cancer* **3**, 89–101 (2003).

38. Hanahan, D. Hallmarks of Cancer: New Dimensions. *Cancer Discov.* **12**, 31–46 (2022).

39. Bostwick, D. G. Grading prostate cancer. *Am. J. Clin. Pathol.* **102**, S38–56 (1994).

40. Bansal, C., Pujani, M., Sharma, K. L., Srivastava, A. N. & Singh, U. S. Grading systems in the cytological diagnosis of breast cancer: a review. *J. Cancer Res. Ther.* **10**, 839–845 (2014).

41. Wang, H. & Unternaehrer, J. J. Epithelial-mesenchymal Transition and Cancer Stem Cells: At the Crossroads of Differentiation and Dedifferentiation. *Dev. Dyn.* **248**, 10–20 (2019).

42. Li, L. & Li, W. Epithelial-mesenchymal transition in human cancer: comprehensive reprogramming of metabolism, epigenetics, and differentiation. *Pharmacol. Ther.* **150**, 33–46 (2015).

43. Asselin-Labat, M.-L. *et al.* Gata-3 is an essential regulator of mammary-gland morphogenesis and luminal-cell differentiation. *Nat. Cell Biol.* **9**, 201–209 (2007).

44. Kouros-Mehr, H., Slorach, E. M., Sternlicht, M. D. & Werb, Z. GATA-3 maintains the differentiation of the luminal cell fate in the mammary gland. *Cell* **127**, 1041–1055 (2006).

45. Kouros-Mehr, H. *et al.* GATA-3 links tumor differentiation and dissemination in a luminal breast cancer model. *Cancer Cell* **13**, 141–152 (2008).

46. Gawrzak, S. *et al.* MSK1 regulates luminal cell differentiation and metastatic dormancy in ER+ breast cancer. *Nat. Cell Biol.* **20**, 211–221 (2018).

47. Elston, C. W. The assessment of histological differentiation in breast cancer. *Aust. N. Z. J. Surg.* **54**, 11–15 (1984).

48. Petushi, S., Garcia, F. U., Haber, M. M., Katsinis, C. & Tozeren, A. Large-scale computations on histology images reveal grade-differentiating parameters for breast cancer. *BMC Med. Imaging* **6**, 14 (2006).

49. Lorentzen, A. *et al.* Expression of NDRG2 is down-regulated in high-risk adenomas and colorectal carcinoma. *BMC Cancer* **7**, 192 (2007).

50. Shen, L. *et al.* NDRG2 facilitates colorectal cancer differentiation through the regulation of

Skp2-p21/p27 axis. *Oncogene* **37**, 1759–1774 (2018).

51. Purdie, C. A. & Piris, J. Histopathological grade, mucinous differentiation and DNA ploidy in relation to prognosis in colorectal carcinoma. *Histopathology* **36**, 121–126 (2000).

52. Qin, J. *et al.* The PSA(-/lo) prostate cancer cell population harbors self-renewing long-term tumor-propagating cells that resist castration. *Cell Stem Cell* **10**, 556–569 (2012).

53. Kim, J. *et al.* FOXA1 inhibits prostate cancer neuroendocrine differentiation. *Oncogene* **36**, 4072–4080 (2017).

54. Chen, W.-Y. *et al.* Androgen deprivation-induced ZBTB46-PTGS1 signaling promotes neuroendocrine differentiation of prostate cancer. *Cancer Lett.* **440-441**, 35–46 (2019).

55. Partin, A. W. *et al.* Prostate specific antigen in the staging of localized prostate cancer: influence of tumor differentiation, tumor volume and benign hyperplasia. *J. Urol.* **143**, 747–752 (1990).

56. Jiang, H.-N. *et al.* Involvement of TRPC channels in lung cancer cell differentiation and the correlation analysis in human non-small cell lung cancer. *PLoS One* **8**, e67637 (2013).

57. Lim, J. S. *et al.* Intratumoural heterogeneity generated by Notch signalling promotes small-cell lung cancer. *Nature* **545**, 360–364 (2017).

58. Hassan, K. A., Chen, G., Kalemkerian, G. P., Wicha, M. S. & Beer, D. G. An Embryonic Stem Cell–Like Signature Identifies Poorly Differentiated Lung Adenocarcinoma but not Squamous Cell Carcinoma. *Clin. Cancer Res.* **15**, 6386–6390 (2009).

59. Wang, B.-Y. *et al.* Lung cancer and prognosis in taiwan: a population-based cancer registry. *J. Thorac. Oncol.* **8**, 1128–1135 (2013).

60. Sun, Z. *et al.* Histologic grade is an independent prognostic factor for survival in non–small cell lung cancer: An analysis of 5018 hospital- and 712 population-based cases. *J. Thorac. Cardiovasc. Surg.* **131**, 1014–1020 (2006).

61. Somnay, Y. R. *et al.* Notch3 expression correlates with thyroid cancer differentiation, induces apoptosis, and predicts disease prognosis. *Cancer* **123**, 769–782 (2017).

62. Ferretti, E. *et al.* Notch signaling is involved in expression of thyrocyte differentiation markers and is

down-regulated in thyroid tumors. *J. Clin. Endocrinol. Metab.* **93**, 4080–4087 (2008).

63. Yu, X.-M., Phan, T., Patel, P. N., Jaskula-Sztul, R. & Chen, H. Chrysin activates Notch1 signaling and suppresses tumor growth of anaplastic thyroid carcinoma in vitro and in vivo. *Cancer* **119**, 774–781 (2013).

64. Akslen, L. A. & LiVolsi, V. A. Prognostic significance of histologic grading compared with subclassification of papillary thyroid carcinoma. *Cancer* **88**, 1902–1908 (2000).

65. Akslen, L. A. Prognostic importance of histologic grading in papillary thyroid carcinoma. *Cancer* **72**, 2680–2685 (1993).

66. Shaha, A. R., Shah, J. P. & Loree, T. R. Patterns of nodal and distant metastasis based on histologic varieties in differentiated carcinoma of the thyroid. *Am. J. Surg.* **172**, 692–694 (1996).

67. Warrick, J. I. *et al.* Intratumoral Heterogeneity of Bladder Cancer by Molecular Subtypes and Histologic Variants. *Eur. Urol.* **75**, 18–22 (2019).

68. Shin, K. *et al.* Hedgehog signaling restrains bladder cancer progression by eliciting stromal production of urothelial differentiation factors. *Cancer Cell* **26**, 521–533 (2014).

69. Wasco, M. J. *et al.* Urothelial carcinoma with divergent histologic differentiation (mixed histologic features) predicts the presence of locally advanced bladder cancer when detected at transurethral resection. *Urology* **70**, 69–74 (2007).

70. Pauwels, R. P., Schapers, R. F., Smeets, A. W., Debruyne, F. M. & Geraedts, J. P. Grading in superficial bladder cancer. (1). Morphological criteria. *Br. J. Urol.* **61**, 129–134 (1988).

71. Hu, S. *et al.* The function of Notch1 intracellular domain in the differentiation of gastric cancer. *Oncol. Lett.* **15**, 6171–6178 (2018).

72. Katz, J. P. *et al.* Loss of Klf4 in mice causes altered proliferation and differentiation and precancerous changes in the adult stomach. *Gastroenterology* **128**, 935–945 (2005).

73. Choe, G., Kim, W. H., Park, J. G. & Kim, Y. I. Effect of suramin on differentiation of human stomach cancer cell lines. *J. Korean Med. Sci.* **12**, 433–442 (1997).

74. Adachi, Y. *et al.* Pathology and prognosis of gastric carcinoma: well versus poorly differentiated

type. *Cancer* **89**, 1418–1424 (2000).

75. Wang, J. & Yue, X. Role and importance of the expression of transcription factor FOXC2 in cervical cancer. *Oncol. Lett.* **14**, 6627–6631 (2017).

76. Wu, X. *et al.* Identification of Key Genes and Pathways in Cervical Cancer by Bioinformatics Analysis. *Int. J. Med. Sci.* **16**, 800–812 (2019).

77. Kidd, E. A. *et al.* Cervical cancer histology and tumor differentiation affect 18F-fluorodeoxyglucose uptake. *Cancer* **115**, 3548–3554 (2009).

78. Anichini, A. *et al.* Skewed T-cell differentiation in patients with indolent non-Hodgkin lymphoma reversed by ex vivo T-cell culture with gammac cytokines. *Blood* **107**, 602–609 (2006).

79. Habeshaw, J. A., Catley, P. F., Stansfeld, A. G. & Brearley, R. L. Surface phenotyping, histology and the nature of non-Hodgkin lymphoma in 157 patients. *Br. J. Cancer* **40**, 11–34 (1979).

80. Seegmiller, A. C., Xu, Y., McKenna, R. W. & Karandikar, N. J. Immunophenotypic differentiation between neoplastic plasma cells in mature B-cell lymphoma vs plasma cell myeloma. *Am. J. Clin. Pathol.* **127**, 176–181 (2007).

81. Micci, F. *et al.* Genomic aberrations in carcinomas of the uterine corpus. *Genes Chromosomes Cancer* **40**, 229–246 (2004).

82. Tafe, L. J., Garg, K., Chew, I., Tornos, C. & Soslow, R. A. Endometrial and ovarian carcinomas with undifferentiated components: clinically aggressive and frequently underrecognized neoplasms. *Mod. Pathol.* **23**, 781–789 (2010).

83. Mo, Z. *et al.* Expression of PD‑1, PD‑L1 and PD‑L2 is associated with differentiation status and histological type of endometrial cancer. *Oncol. Lett.* **12**, 944–950 (2016).

84. Creasman, W. T. Prognostic significance of hormone receptors in endometrial cancer. *Cancer* **71**, 1467–1470 (1993).

85. Liu, G. J. *et al.* Pax5 loss imposes a reversible differentiation block in B-progenitor acute lymphoblastic leukemia. *Genes Dev.* **28**, 1337–1350 (2014).

86. Lee, E. J., Pollak, A., Leavitt, R. D., Testa, J. R. & Schiffer, C. A. Minimally differentiated acute

nonlymphocytic leukemia: a distinct entity. *Blood* **70**, 1400–1406 (1987).

87. Mueller, B. U. *et al.* ATRA resolves the differentiation block in t(15;17) acute myeloid leukemia by restoring PU.1 expression. *Blood* **107**, 3330–3338 (2006).

88. Que, W.-C., Qiu, H.-Q., Cheng, Y., Liu, M.-B. & Wu, C.-Y. PTEN in kidney cancer: A review and meta-analysis. *Clin. Chim. Acta* **480**, 92–98 (2018).

89. Leibovich, B. C. *et al.* Histological subtype is an independent predictor of outcome for patients with renal cell carcinoma. *J. Urol.* **183**, 1309–1315 (2010).

90. Prasad, S. R. *et al.* Common and uncommon histologic subtypes of renal cell carcinoma: imaging spectrum with pathologic correlation. *Radiographics* **26**, 1795–806; discussion 1806–10 (2006).

91. Ekmekcioglu, S. *et al.* Down-regulated melanoma differentiation associated gene (mda-7) expression in human melanomas. *International journal of cancer* **94**, 54–59 (2001).

92. Leung, K.-W. *et al.* Pin1 overexpression is associated with poor differentiation and survival in oral squamous cell carcinoma. *Oncol. Rep.* **21**, 1097–1104 (2009).

93. Ogden, G. R., Chisholm, D. M., Adi, M. & Lane, E. B. Cytokeratin expression in oral cancer and its relationship to tumor differentiation. *J. Oral Pathol. Med.* **22**, 82–86 (1993).

94. Warnakulasuriya, S. Histological grading of oral epithelial dysplasia: revisited. *J. Pathol.* **194**, 294–297 (2001).

95. Strieder, L. *et al.* Comparative analysis of three histologic grading methods for squamous cell carcinoma of the lip. *Oral Dis.* **23**, 120–125 (2017).

96. Guichet, P.-O. *et al.* Cell death and neuronal differentiation of glioblastoma stem-like cells induced by neurogenic transcription factors. *Glia* **61**, 225–239 (2013).

97. Boso, D. *et al.* HIF-1α/Wnt signaling-dependent control of gene transcription regulates neuronal differentiation of glioblastoma stem cells. *Theranostics* **9**, 4860–4877 (2019).

98. Rampazzo, E. *et al.* Wnt activation promotes neuronal differentiation of glioblastoma. *Cell Death Dis.* **4**, e500 (2013).

99. Zhang, Q. B. *et al.* Differentiation profile of brain tumor stem cells: a comparative study with neural

stem cells. *Cell Res.* **16**, 909–915 (2006).

100. Ricci-Vitiani, L. *et al.* Tumour vascularization via endothelial differentiation of glioblastoma stem-like cells. *Nature* **468**, 824–828 (2010).

101. Ricci-Vitiani, L. *et al.* Mesenchymal differentiation of glioblastoma stem cells. *Cell Death Differ.* **15**, 1491–1498 (2008).

102. Rose, S. L. Notch signaling pathway in ovarian cancer. *Int. J. Gynecol. Cancer* **19**, 564–566 (2009).

103. McAuliffe, S. M. *et al.* Targeting Notch, a key pathway for ovarian cancer stem cells, sensitizes tumors to platinum therapy. *Proc. Natl. Acad. Sci. U. S. A.* **109**, E2939–48 (2012).

104. Rose, S. L., Kunnimalaiyaan, M., Drenzek, J. & Seiler, N. Notch 1 signaling is active in ovarian cancer. *Gynecol. Oncol.* **117**, 130–133 (2010).

105. Wang, M., Wang, J., Wang, L., Wu, L. & Xin, X. Notch1 expression correlates with tumor differentiation status in ovarian carcinoma. *Med. Oncol.* **27**, 1329–1335 (2010).

106. Malpica, A. Grading of ovarian cancer: a histotype-specific approach. *Int. J. Gynecol. Pathol.* **27**, 175–181 (2008).

107. Silverberg, S. G. Histopathologic grading of ovarian carcinoma: a review and proposal. *Int. J. Gynecol. Pathol.* **19**, 7–15 (2000).

108. Shachaf, C. M. *et al.* MYC inactivation uncovers pluripotent differentiation and tumour dormancy in hepatocellular cancer. *Nature* **431**, 1112–1117 (2004).

109. Fang, S. *et al.* Lymphoid enhancer-binding factor-1 promotes stemness and poor differentiation of hepatocellular carcinoma by directly activating the NOTCH pathway. *Oncogene* **38**, 4061–4074 (2019).

110. Jang, H.-J., Kim, T. K., Burns, P. N. & Wilson, S. R. Enhancement patterns of hepatocellular carcinoma at contrast-enhanced US: comparison with histologic differentiation. *Radiology* **244**, 898–906 (2007).

111. Calderaro, J., Ziol, M., Paradis, V. & Zucman-Rossi, J. Molecular and histological correlations in liver cancer. *J. Hepatol.* **71**, 616–630 (2019).

112. Gao, Y.-B. *et al.* Genetic landscape of esophageal squamous cell carcinoma. *Nat. Genet.* **46**, 1097–1102 (2014).

113. Natsuizaka, M. *et al.* Interplay between Notch1 and Notch3 promotes EMT and tumor initiation in squamous cell carcinoma. *Nat. Commun.* **8**, 1758 (2017).

114. Trivers, K. F., Sabatino, S. A. & Stewart, S. L. Trends in esophageal cancer incidence by histology, United States, 1998-2003. *Int. J. Cancer* **123**, 1422–1428 (2008).

115. Nadal, A. & Cardesa, A. Molecular biology of laryngeal squamous cell carcinoma. *Virchows Arch.* **442**, 1–7 (2003).

116. Passler, C. *et al.* Anaplastic (undifferentiated) thyroid carcinoma (ATC). *Langenbecks. Arch. Surg.* **384**, 284–293 (1999).

117. Sarioglu, S. *et al.* Undifferentiated Laryngeal Carcinoma with Pagetoid Spread. *Head Neck Pathol.* **10**, 252–255 (2016).

118. Stanley, R. J., Weiland, L. H., DeSanto, L. W. & Neel, H. B., 3rd. Lymphoepithelioma (undifferentiated carcinoma) of the laryngohypopharynx. *Laryngoscope* **95**, 1077–1081 (1985).

119. Matsui, W. *et al.* Clonogenic multiple myeloma progenitors, stem cell properties, and drug resistance. *Cancer Res.* **68**, 190–197 (2008).

120. Billadeau, D., Ahmann, G., Greipp, P. & Van Ness, B. The bone marrow of multiple myeloma patients contains B cell populations at different stages of differentiation that are clonally related to the malignant plasma cell. *J. Exp. Med.* **178**, 1023–1031 (1993).

121. Subramanian, R., Basu, D. & Dutta, T. K. Prognostic significance of bone marrow histology in multiple myeloma. *Indian J. Cancer* **46**, 40–45 (2009).

122. Bartl, R. *et al.* Histologic classification and staging of multiple myeloma. A retrospective and prospective study of 674 cases. *Am. J. Clin. Pathol.* **87**, 342–355 (1987).

123. Cancer today. https://gco.iarc.fr/today/online-analysis-multi-bars?v=2018&mode=cancer&mode_population=countries&population=900&populations=900&key=total&sex=0&cancer=39&type=0&statistic=5&preval

ence=0&population_group=0&ages_group%5B%5D=0&ages_group%5B%5D=17&nb_items=55&group_cancer=1&include_nmsc=1&include_nmsc_other=1&type_multiple=%257B%2522inc%2522%253Afalse%252C%2522mort%2522%253Atrue%252C%2522prev%2522%253Afalse%257D&orientation=horizontal&type_sort=0&type_nb_items=%257B%2522top%2522%253Atrue%252C%2522bottom%2522%253Afalse%257D&population_group_globocan_id=.

124. Zhang, X., Cruz, F. D., Terry, M., Remotti, F. & Matushansky, I. Terminal differentiation and loss of tumorigenicity of human cancers via pluripotency-based reprogramming. *Oncogene* **32**, 2249–60, 2260.e1–21 (2013).

125. Cairns, J. Mutation selection and the natural history of cancer. *Nature* **255**, 197–200 (1975).

126. Snippert, H. J. *et al.* Intestinal crypt homeostasis results from neutral competition between symmetrically dividing Lgr5 stem cells. *Cell* **143**, 134–144 (2010).

127. Clevers, H. The intestinal crypt, a prototype stem cell compartment. *Cell* **154**, 274–284 (2013).

128. van der Heijden, M. & Vermeulen, L. Stem cells in homeostasis and cancer of the gut. *Mol. Cancer* **18**, 66 (2019).

129. Gehart, H. & Clevers, H. Tales from the crypt: new insights into intestinal stem cells. *Nat. Rev. Gastroenterol. Hepatol.* **16**, 19–34 (2019).

130. Chandramouly, G., Abad, P. C., Knowles, D. W. & Lelièvre, S. A. The control of tissue architecture over nuclear organization is crucial for epithelial cell fate. *J. Cell Sci.* **120**, 1596–1606 (2007).

131. Sprouffske, K., Pepper, J. W. & Maley, C. C. Accurate reconstruction of the temporal order of mutations in neoplastic progression. *Cancer Prev. Res.* **4**, 1135–1144 (2011).

132. Haeno, H., Levine, R. L., Gilliland, D. G. & Michor, F. A progenitor cell origin of myeloid malignancies. *Proc. Natl. Acad. Sci. U. S. A.* **106**, 16616–16621 (2009).

133. Sprouffske, K. *et al.* An evolutionary explanation for the presence of cancer nonstem cells in neoplasms. *Evol. Appl.* **6**, 92–101 (2013).

134. Min, B. Spontaneous T Cell Proliferation: A Physiologic Process to Create and Maintain Homeostatic Balance and Diversity of the Immune System. *Front. Immunol.* **9**, 547 (2018).

135. Liu, W.-H. *et al.* Hepatocyte proliferation during liver regeneration is impaired in mice with methionine diet-induced hyperhomocysteinemia. *Am. J. Pathol.* **177**, 2357–2365 (2010).

136. Manohar, R. & Lagasse, E. Chapter 45 - Liver Stem Cells. in *Principles of Tissue Engineering (Fourth Edition)* (eds. Lanza, R., Langer, R. & Vacanti, J.) 935–950 (Academic Press, 2014).

137. Zhong, F. & Jiang, Y. Endogenous Pancreatic β Cell Regeneration: A Potential Strategy for the Recovery of β Cell Deficiency in Diabetes. *Front. Endocrinol.* **10**, 101 (2019).

138. Luckheeram, R. V., Zhou, R., Verma, A. D. & Xia, B. CD4$^+$T cells: differentiation and functions. *Clin. Dev. Immunol.* **2012**, 925135 (2012).

139. Grotendorst, G. R., Rahmanie, H. & Duncan, M. R. Combinatorial signaling pathways determine fibroblast proliferation and myofibroblast differentiation. *FASEB J.* **18**, 469–479 (2004).

140. Visco, V. *et al.* Human colon fibroblasts induce differentiation and proliferation of intestinal epithelial cells through the direct paracrine action of keratinocyte growth factor. *J. Cell. Physiol.* **220**, 204–213 (2009).

141. Sriuranpong, V. *et al.* Notch signaling induces cell cycle arrest in small cell lung cancer cells. *Cancer Res.* **61**, 3200–3205 (2001).

142. Sjölund, J., Manetopoulos, C., Stockhausen, M.-T. & Axelson, H. The Notch pathway in cancer: differentiation gone awry. *Eur. J. Cancer* **41**, 2620–2629 (2005).

143. Weng, A. P. *et al.* Activating mutations of NOTCH1 in human T cell acute lymphoblastic leukemia. *Science* **306**, 269–271 (2004).

144. O'Neil, J. *et al.* Activating Notch1 mutations in mouse models of T-ALL. *Blood* **107**, 781–785 (2006).

145. Ferrando, A. A. The role of NOTCH1 signaling in T-ALL. *Hematology Am. Soc. Hematol. Educ. Program* 353–361 (2009).

146. Vilimas, T. *et al.* Targeting the NF-kappaB signaling pathway in Notch1-induced T-cell leukemia. *Nat. Med.* **13**, 70–77 (2007).

147. Sulis, M. L. *et al.* NOTCH1 extracellular juxtamembrane expansion mutations in T-ALL. *Blood* **112**,

733–740 (2008).

148. Cullion, K. *et al.* Targeting the Notch1 and mTOR pathways in a mouse T-ALL model. *Blood* **113**, 6172–6181 (2009).

149. Rangarajan, A. *et al.* Notch signaling is a direct determinant of keratinocyte growth arrest and entry into differentiation. *EMBO J.* **20**, 3427–3436 (2001).

150. Cole, M. D. The myc oncogene: its role in transformation and differentiation. *Annu. Rev. Genet.* **20**, 361–384 (1986).

151. Coppola, J. A. & Cole, M. D. Constitutive c-myc oncogene expression blocks mouse erythroleukaemia cell differentiation but not commitment. *Nature* **320**, 760–763 (1986).

152. Freytag, S. O. Enforced expression of the c-myc oncogene inhibits cell differentiation by precluding entry into a distinct predifferentiation state in G0/G1. *Mol. Cell. Biol.* **8**, 1614–1624 (1988).

153. Wilson, A. *et al.* c-Myc controls the balance between hematopoietic stem cell self-renewal and differentiation. *Genes Dev.* **18**, 2747–2763 (2004).

154. Sun, C. *et al.* TMPRSS2-ERG fusion, a common genomic alteration in prostate cancer activates C-MYC and abrogates prostate epithelial differentiation. *Oncogene* **27**, 5348–5353 (2008).

155. Pelengaris, S. & Khan, M. The many faces of c-MYC. *Arch. Biochem. Biophys.* **416**, 129–136 (2003).

156. Prochownik, E. V. & Kukowska, J. Deregulated expression of c-myc by murine erythroleukaemia cells prevents differentiation. *Nature* **322**, 848–850 (1986).

157. Dmitrovsky, E. *et al.* Expression of a transfected human c-myc oncogene inhibits differentiation of a mouse erythroleukaemia cell line. *Nature* **322**, 748–750 (1986).

158. Gandarillas, A. & Watt, F. M. c-Myc promotes differentiation of human epidermal stem cells. *Genes Dev.* **11**, 2869–2882 (1997).

159. Klinakis, A. *et al.* A novel tumour-suppressor function for the Notch pathway in myeloid leukaemia. *Nature* **473**, 230–233 (2011).

160. Watt, F. M., Estrach, S. & Ambler, C. A. Epidermal Notch signalling: differentiation, cancer and

adhesion. *Curr. Opin. Cell Biol.* **20**, 171–179 (2008).

161. Panelos, J. & Massi, D. Emerging role of Notch signaling in epidermal differentiation and skin cancer. *Cancer Biol. Ther.* **8**, 1986–1993 (2009).

162. He, S. *et al.* Expression of Lgr5, a marker of intestinal stem cells, in colorectal cancer and its clinicopathological significance. *Biomed. Pharmacother.* **68**, 507–513 (2014).

163. Wu, X.-S., Xi, H.-Q. & Chen, L. Lgr5 is a potential marker of colorectal carcinoma stem cells that correlates with patient survival. *World J. Surg. Oncol.* **10**, 244 (2012).

164. Herbst, A. *et al.* Comprehensive analysis of β-catenin target genes in colorectal carcinoma cell lines with deregulated Wnt/β-catenin signaling. *BMC Genomics* **15**, 74 (2014).

165. Enane, F. O. *et al.* GATA4 loss of function in liver cancer impedes precursor to hepatocyte transition. *J. Clin. Invest.* **127**, 3527–3542 (2017).

166. Yu, X.-M. *et al.* Notch1 Signaling Regulates the Aggressiveness of Differentiated Thyroid Cancer and Inhibits SERPINE1 Expression. *Clin. Cancer Res.* **22**, 3582–3592 (2016).

167. Cancer Genome Atlas Research Network. Integrated genomic characterization of papillary thyroid carcinoma. *Cell* **159**, 676–690 (2014).

168. Busto Catañón, L., Sánchez Merino, J. M., Picallo Sánchez, J. A. & Gelabert Mas, A. [Clinical prognostic factors in superficial cancer of the urinary bladder]. *Arch. Esp. Urol.* **54**, 131–138 (2001).

169. Nishida, T., Katayama, S., Tsujimoto, M., Nakamura, J. & Matsuda, H. Clinicopathological significance of poorly differentiated thyroid carcinoma. *Am. J. Surg. Pathol.* **23**, 205–211 (1999).

170. Schwede, M. *et al.* Stem cell-like gene expression in ovarian cancer predicts type II subtype and prognosis. *PLoS One* **8**, e57799 (2013).

171. Smith, B. A. *et al.* A basal stem cell signature identifies aggressive prostate cancer phenotypes. *Proc. Natl. Acad. Sci. U. S. A.* **112**, E6544–52 (2015).

172. Toraih, E. A. *et al.* Stemness-related transcriptional factors and homing gene expression profiles in hepatic differentiation and cancer. *Mol. Med.* **22**, 653–663 (2016).

173. Riester, M. *et al.* Distance in cancer gene expression from stem cells predicts patient survival. *PLoS*

*One* **12**, e0173589 (2017).

174. Mueller, E. *et al.* Terminal Differentiation of Human Breast Cancer through PPARγ. *Mol. Cell* **1**, 465–470 (1998).

175. Sarraf, P. *et al.* Differentiation and reversal of malignant changes in colon cancer through PPARγ. *Nat. Med.* **4**, 1046–1052 (1998).

176. Dow, L. E. *et al.* Apc Restoration Promotes Cellular Differentiation and Reestablishes Crypt Homeostasis in Colorectal Cancer. *Cell* **161**, 1539–1552 (2015).

177. Pham, P. V. *et al.* Differentiation of breast cancer stem cells by knockdown of CD44: promising differentiation therapy. *J. Transl. Med.* **9**, 209 (2011).

178. Wang, Z.-Y. & Chen, Z. Acute promyelocytic leukemia: from highly fatal to highly curable. *Blood* **111**, 2505–2515 (2008).

179. Christian, S. *et al.* The novel dihydroorotate dehydrogenase (DHODH) inhibitor BAY 2402234 triggers differentiation and is effective in the treatment of myeloid malignancies. *Leukemia* (2019) doi:10.1038/s41375-019-0461-5.

180. Petrie, K., Zelent, A. & Waxman, S. Differentiation therapy of acute myeloid leukemia: past, present and future. *Curr. Opin. Hematol.* **16**, 84–91 (2009).

181. Johnson, D. E. & Redner, R. L. An ATRActive future for differentiation therapy in AML. *Blood Rev.* **29**, 263–268 (2015).

182. Ferrara, F. F. *et al.* Histone deacetylase-targeted treatment restores retinoic acid signaling and differentiation in acute myeloid leukemia. *Cancer Res.* **61**, 2–7 (2001).

183. Takebe, N. & Ivy, S. P. Controversies in cancer stem cells: targeting embryonic signaling pathways. *Clin. Cancer Res.* **16**, 3106–3112 (2010).

184. Takebe, N., Harris, P. J., Warren, R. Q. & Ivy, S. P. Targeting cancer stem cells by inhibiting Wnt, Notch, and Hedgehog pathways. *Nat. Rev. Clin. Oncol.* **8**, 97–106 (2011).

185. Hu, Y. & Fu, L. Targeting cancer stem cells: a new therapy to cure cancer patients. *Am. J. Cancer Res.* **2**, 340 (2012).

186. Massard, C., Deutsch, E. & Soria, J.-C. Tumour stem cell-targeted treatment: elimination or differentiation. *Ann. Oncol.* **17**, 1620–1624 (2006).

187. Frank, N. Y., Schatton, T. & Frank, M. H. The therapeutic promise of the cancer stem cell concept. *J. Clin. Invest.* **120**, 41–50 (2010).

188. Sell, S. Cancer stem cells and differentiation therapy. *Tumour Biol.* **27**, 59–70 (2006).

189. Sell, S. Stem cell origin of cancer and differentiation therapy. *Crit. Rev. Oncol. Hematol.* **51**, 1–28 (2004).

190. Lapidot, T. *et al.* A cell initiating human acute myeloid leukaemia after transplantation into SCID mice. *Nature* **367**, 645–648 (1994).

191. Visvader, J. E. & Lindeman, G. J. Cancer stem cells in solid tumours: accumulating evidence and unresolved questions. *Nat. Rev. Cancer* **8**, 755–768 (2008).

192. Todaro, M. *et al.* Colon cancer stem cells dictate tumor growth and resist cell death by production of interleukin-4. *Cell Stem Cell* **1**, 389–402 (2007).

193. Lombardo, Y. *et al.* Bone morphogenetic protein 4 induces differentiation of colorectal cancer stem cells and increases their response to chemotherapy in mice. *Gastroenterology* **140**, 297–309 (2011).

194. Yan, Y. *et al.* All-trans retinoic acids induce differentiation and sensitize a radioresistant breast cancer cells to chemotherapy. *BMC Complement. Altern. Med.* **16**, 113 (2016).

195. Brodeur, G. M. Spontaneous regression of neuroblastoma. *Cell Tissue Res.* **372**, 277–286 (2018).

196. van Groningen, T. *et al.* Neuroblastoma is composed of two super-enhancer-associated differentiation states. *Nat. Genet.* **49**, 1261–1266 (2017).

197. Reynolds, C. P., Matthay, K. K., Villablanca, J. G. & Maurer, B. J. Retinoid therapy of high-risk neuroblastoma. *Cancer Lett.* **197**, 185–192 (2003).

198. Reynolds, C. P. Differentiating agents in pediatric malignancies: retinoids in neuroblastoma. *Curr. Oncol. Rep.* **2**, 511–518 (2000).